\documentclass{PoS}
\usepackage{epsfig}

\PoS{PoS(BDMH2004)}

\title{The galaxy-dark matter connection}

\ShortTitle{The galaxy-dark matter connection}

\author{\speaker{Frank C. van den Bosch}\\
        Department of Physics, ETH Zürich, Switzerland\\
        E-mail: \email{vdbosch@phys.ethz.ch}}

\author{Xiaohu Yang and H.J. Mo\\
        Department of Astronomy, University of Massachusetts, USA}
      
      \abstract{What galaxy  lives in what  halo?  The answer  to this
        simple question  holds important information  regarding galaxy
        formation  and  evolution.   We  describe  a  new  statistical
        technique  to link galaxies  to their  dark matter  haloes, or
        light to mass, using  the clustering properties of galaxies as
        function   of  their   luminosity.   The   galaxy-dark  matter
        connection  thus established,  and  parameterized through  the
        conditional luminosity function, indicates the presence of two
        characteristic  scales  in  galaxy  formation:  one  at  $\sim
        10^{11}  h^{-1}   \Msun$,  where  galaxy   formation  is  most
        efficient, and another at $\sim 10^{13} h^{-1} \Msun$, where a
        transition  occurs from  systems dominated  by  one brightest,
        central galaxy  to systems  with several dominant  galaxies of
        comparable luminosity.  We test the relation between light and
        mass established  from galaxy clustering  alone with dynamical
        masses obtained from satellite  kinematics, and show that both
        are   in  excellent   agreement.   We   also  present   a  new
        (halo-based)   galaxy-group   finder,   and  show   that   the
        multiplicity  function  of  galaxy  groups identified  in  the
        2dFGRS suggests  a relatively high mass-to-light  ratio on the
        scales of galaxy clusters, or, alternatively, a relatively low
        value of  the power-spectrum normalization  $\sigma_8$.  These
        findings  are  also  supported  by  our  studies  of  pairwise
        peculiar  velocities and  satellite  abundances.  Finally,  we
        directly  measure  the  halo  occupation statistics  from  our
        galaxy groups, which  are a good proxy of  dark matter haloes,
        and  show  that these  are  in  excellent  agreement with  our
        conditional luminosity function model.}

\FullConference{Baryons in Dark Matter Halos\\
                 5-9 October 2004\\
                 Novigrad, Croatia}

\newcommand{\etal}{{et al.~}}

\newcommand{\Msun}{\>{\rm M_{\odot}}}

\newcommand{\MLsun}{\>({\rm M}/{\rm L})_{\odot}}

\newcommand{\apj}{ApJ}

\newcommand{\mnras}{MNRAS}

\def\gtsima{$\; \buildrel > \over \sim \;$}
\def\ltsima{$\; \buildrel < \over \sim \;$}
\def\prosima{$\; \buildrel \propto \over \sim \;$}
\def\gsim{\lower.7ex\hbox{\gtsima}}
\def\lsim{\lower.7ex\hbox{\ltsima}}
\def\simgt{\lower.7ex\hbox{\gtsima}}
\def\simlt{\lower.7ex\hbox{\ltsima}}
\def\simpr{\lower.7ex\hbox{\prosima}}
\def\la{\lsim}
\def\ga{\gsim}
\def\lta{\la}
\def\gta{\ga}

\begin{document}

\section{Introduction}
\label{sec:intro}

According  to the  current paradigm  of structure  formation, galaxies
form and reside inside extended cold dark matter (CDM) haloes.  One of
the  ultimate challenges in  astrophysics, therefore,  is to  obtain a
detailed  understanding  of   how  galaxies  with  different  physical
properties occupy  dark matter haloes of different  masses.  This link
between  galaxies and  dark matter  haloes  is an  imprint of  various
complicated physical  processes related  to galaxy formation,  such as
cooling, star formation, merging, tidal interactions, and a variety of
feedback  processes.  Although  the statistical  link itself  does not
give  a physical  explanation  of  how galaxies  form  and evolve,  it
provides  important constraints on  these processes  and on  how their
efficiencies scale with halo mass.

To  quantify  the  relationship  between  haloes  and  galaxies  in  a
statistical way, it has become customary to specify the so-called halo
occupation distribution,  $P(N \vert M)$, which  gives the probability
to find  $N$ galaxies  (with some specified  properties) in a  halo of
mass $M$.  This occupation  distribution can be constrained using data
on the  clustering properties of galaxies, as  it completely specifies
the  galaxy  bias, and  has  been  used  extensively to  study  galaxy
occupation statistics  and large  scale structure (e.g.,  [2,9,18] and
references therein).

\section{The Conditional Luminosity Function}
\label{sec:clf}

Arguably, one of the most important physical properties of a galaxy is
its total luminosity. Ideally, one would therefore consider occupation
statistics  as a  function of  luminosity. In  particular,  this would
allow  a  direct  estimate  of  the average  mass-to-light  ratios  as
function of  halo mass, as well  as the construction  of mock redshift
surveys.   In  a  series  of  papers [12,18],  we  therefore  included
luminosities  in the  halo  occupation statistics  by introducing  the
conditional  luminosity function  (CLF)  $\Phi(L \vert  M) {\rm  d}L$,
which gives  the average number  of galaxies with luminosities  in the
range $L \pm {\rm  d}L/2$ that reside in a halo of  mass $M$.  The CLF
is the  direct link between  the galaxy luminosity  function $\Phi(L)$
and the halo mass function $n(M)$, according to
\begin{equation}
\label{phiL}
\Phi(L) = \int_{0}^{\infty} \Phi(L \vert M) \, n(M) \, {\rm d}M.
\end{equation}
In CDM  cosmologies, more massive  haloes are more  strongly clustered
(e.g.,  [6]).  This  means that  information regarding  the clustering
strength  of galaxies  (of  a given  luminosity) contains  information
about the  characteristic mass of the  halo in which  they reside.  At
sufficiently  large   separations,  $r$,  the   two-point  correlation
function of galaxies of luminosity  $L$ is given by $\xi_{\rm gg}(r,L)
= \bar{b}^2(L) \, \xi_{\rm dm}(r)$. Here $\xi_{\rm dm}(r)$ is the dark
matter mass correlation function, and $\bar{b}(L)$ is the average {\it
  bias}  of galaxies  of luminosity  $L$, which  derives from  the CLF
according to
\begin{equation}
\label{biasmod}
\bar{b}(L) = {1 \over \Phi(L)} \int_{0}^{\infty} \Phi(L
\vert M) \, b(M) \, n(M) \, {\rm d}M.
\end{equation}
with $b(M)$  the bias of dark  matter haloes of  mass $M$.  Therefore,
the combination  of an  observed luminosity function,  $\Phi(L)$, plus
measurements of the  galaxy-galaxy two-point correlation function {\it
  as  function of  luminosity} puts  stringent constraints  on $\Phi(L
\vert M)$.
\begin{figure}
\centerline{\epsfig{figure=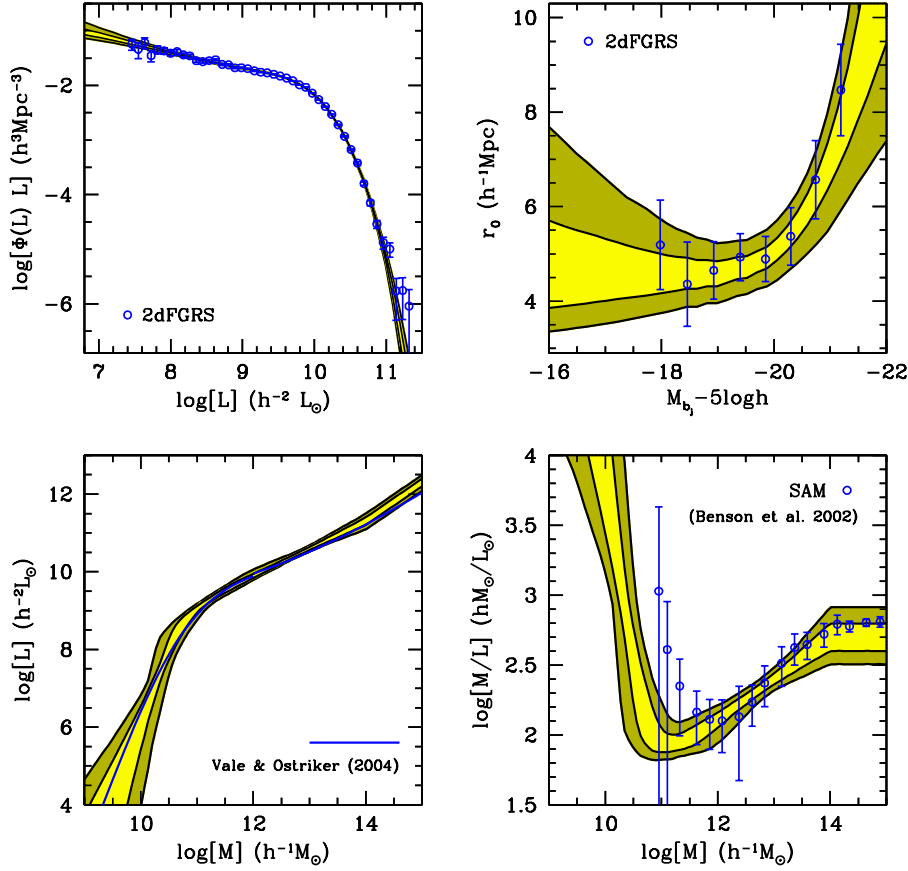,width=0.8\textwidth}}
\caption{Posterior constraints on a number of quantities computed from
  the MCMC described in the text.  The contours show the 68\% and 99\%
  confidence  limits from the  marginalized distribution.   {\it Upper
    left-hand  panel:} The  galaxy luminosity  function;  open circles
  with errorbars correspond  to the 2dFGRS data from  [5].  {\it Upper
    right-hand panel:}  galaxy-galaxy correlation lengths  as function
  of absolute magnitude; open circles with errorbars correspond to the
  2dFGRS  data  from [8].   {\it  Lower  left-hand  panel:} the  total
  luminosity per halo as function of halo mass. Solid line corresponds
  to  the model  of [11],  and is  shown for  comparison.   {\it Lower
    right-hand panel:} the average  mass-to-light ratio as function of
  halo  mass.    Open  circles   with  errorbars  correspond   to  the
  semi-analytical model of  [1] and is shown for  comparison. See text
  for details.}
\label{fig:mcm}
\end{figure}

We assume that  the CLF can be described by  a Schechter function, and
describe the mass dependencies using a total of 8 free parameters (see
[12,15,18]).  We  use a Monte-Carlo  Markov Chain (hereafter  MCMC) to
probe  the  likelihood  function  in our  multi-dimensional  parameter
space, and  to put confidence  levels on all derived  quantities.  The
results obtained for a $\Lambda$CDM `concordance' cosmology ($\Omega_m
= 0.3$, $\Omega_{\Lambda}=0.7$,  $h=0.7$, $\sigma_8=0.9$) are shown in
Fig.~1.  The open  circles with errorbars in the  upper panels are the
data used to  constrain the models.  The shaded  areas indicate the 68
and 99  percent confidence levels  on $\Phi(L)$ and  $r_0(L)$ computed
from the MCMC.  Note the good agreement with the data, indicating that
the CLF  can accurately match  the observed abundances  and clustering
properties of galaxies in the 2dFGRS.  We emphasize that this is not a
trivial  result,  as  the  data  can  only be  fitted  for  a  certain
combination of cosmological parameters (see [13]).

The lower  left-hand panel of  Fig.~1 plots the relation  between halo
mass $M$ and the total halo luminosity $L$, which follows from the CLF
according to
\begin{equation}
\label{LofM}
\langle L \rangle(M) = \int_{0}^{\infty} \Phi(L  \vert M) \, L \, {\rm
d}L
\end{equation}
Note that  the confidence levels  are extremely tight,  especially for
the more massive haloes.  The $L(M)$ relation reveals a dramatic break
at around $M \simeq 10^{11} h^{-1} \Msun$, indicating a characteristic
scale in  galaxy formation.  These results are  in excellent agreement
with the $L(M)$ relation of Vale \& Ostriker ([11]), obtained assuming
a monotonic relation between  light and mass.  This agreement combined
with  the extremely  tight  confidence levels  obtained  from our  CLF
analysis suggests that we have established a robust connection between
light and mass.

Finally, the lower right-hand  panel of Fig.~1 plots the corresponding
mass-to-light ratios as function of halo mass.  The pronounced minimum
in  $\langle M/L \rangle_M$  indicates that  galaxy formation  is most
efficient in haloes with masses  in the range $5 \times 10^{10} h^{-1}
\Msun \lta  M \lta  10^{12} h^{-1} \Msun$.   For less  massive haloes,
$\langle  M/L \rangle_M$  increases drastically  with  decreasing halo
mass, which is required in order  to bring the steep slope of the halo
mass  function at  low $M$  in agreement  with the  relatively shallow
faint-end  slope  of  the  observed  LF.   It  indicates  that  galaxy
formation needs to become extremely inefficient in haloes with $M \lta
5 \times 10^{10} h^{-1} \Msun$ in order to prevent an overabundance of
faint galaxies.  The increase in  $\langle M/L \rangle_M$ from $M \sim
10^{11} h^{-1} \Msun$  to $M \sim 10^{14} h^{-1}  \Msun$ is thought to
be  associated with the  decreasing ability  of the  gas to  cool with
increasing halo  mass.  The open circles with  errorbars correspond to
the semi-analytical model of Benson  \etal ([1]), which has been tuned
to match  the galaxy luminosity function.  It  is extremely reassuring
that  two completely independent  methods yield  average mass-to-light
ratios that are in such good agreement.

Because it  gives a statistical description of  the galaxy-dark matter
connection, the  CLF is  an extremely powerful  tool.  In a  series of
papers we have  used it to investigate large  scale structure [16,19],
the environment dependence of  the galaxy luminosity function [7], the
kinematics and  abundances of satellite galaxies  [14,15], and various
properties of galaxy groups [20,21,22].  In addition, we have used the
CLF  formalism  to  constrain  cosmological  parameters  [13]  and  to
construct detailed mock redshift surveys  [19].  Here we summary a few
of the highlights.

\section{Mock Galaxy Redshift Surveys}
\label{sec:mock}

An important application of the CLF is the construction of mock galaxy
redshift surveys (hereafter MGRSs), which are extremely useful tools to
aid  in the  interpretation of  large redshift  surveys.  As  with any
data-set,  several  observational   biases  hamper  a  straightforward
interpretation of  such surveys (e.g.,  Malmquist bias, redshift-space
distortions, fiber collisions). The CLF  is ideally suited to build up
``virtual'' Universes  from which mock galaxy redshift  surveys can be
constructed using the same biases and incompleteness effects as in the
real data. In [19] we have used the CLF to populate dark matter haloes
in numerical simulations with  galaxies of different luminosities, and
constructed detailed MGRSs that can  be compared on a one-to-one basis
with the 2dFGRS (see also [14,15,20]).
\begin{figure}
  \centerline{\epsfig{figure=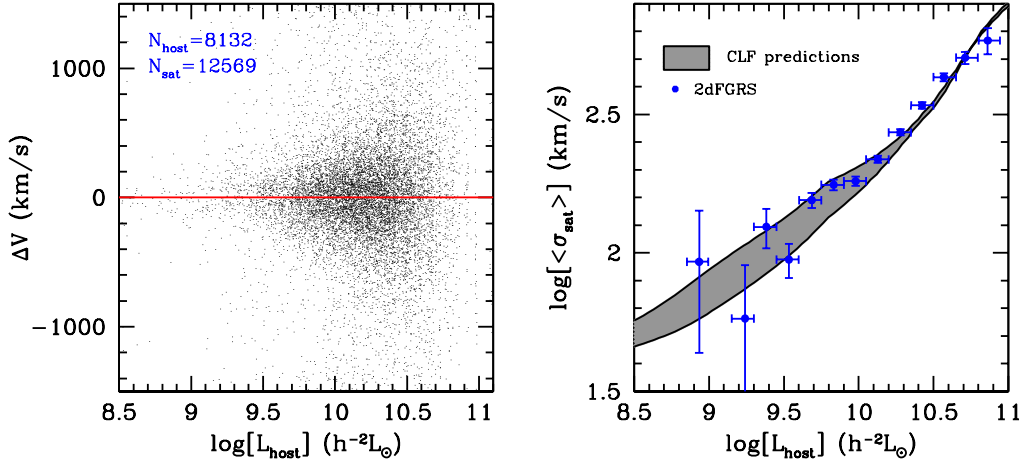,width=0.9\textwidth}}
\caption{{\it Left-hand panel} The difference in line-of-sight velocity, 
  $\Delta V$, between  hosts and satellites in the  2dFGRS as function
  of  the  luminosity  $L_{\rm   host}$  of  the  host  galaxy.   {\it
    Right-hand panel:} Solid dots with errorbars indicate the best-fit
  $\sigma_{\rm sat}(L_{\rm host})$ obtained from the 2dFGRS data shown
  in  the left-hand panel.   The gray  area indicates  the expectation
  values  obtained using the  CLF.  The  excellent agreement  with the
  2dFGRS results  gives an independent, dynamical  confirmation of the
  relation between light and  mass inferred from the galaxy clustering
  properties.}
\label{fig:sats}
\end{figure}

\section{Satellite Kinematics}
\label{sec:sats}

In order  to test  the relation  between light and  mass shown  in the
lower left panel of Fig.~1, and which has been derived solely from the
galaxy  clustering  properties, we  use  the  kinematics of  satellite
galaxies. Since satellites  probe the potential well out  to the outer
edges of their haloes, they  are ideally suited to measure {\it total}
halo  masses  (unlike,  for  example,  a  disk  rotation  curve  which
typically only  probes the potential out  to a fraction  of the virial
radius).  A downside of using satellite galaxies, however, is that the
number  of detectable  satellites in  individual systems  is generally
small. One therefore typically  stacks the data on many host-satellite
pairs to obtain  statistical estimates of halo masses  (see [3] for an
up-to-date review).
\begin{figure}
\centerline{\epsfig{figure=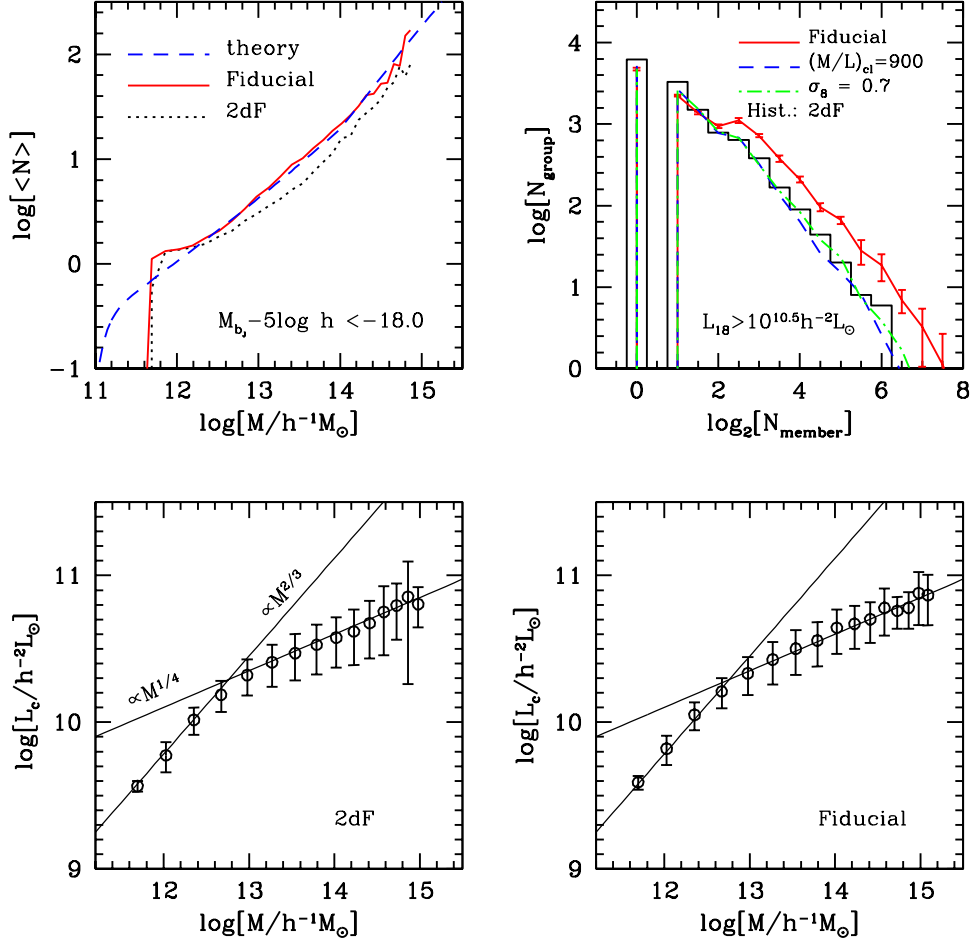,width=0.85\textwidth}}
\caption{{\it Upper left panel:} Mean group occupation numbers
  as function  of halo  mass.  Dashed line  indicates the  `input', as
  specified by the  CLF used to construct the  MGRS.  Solid and dotted
  lines indicate  results obtained from the groups  extracted with our
  halo-based group finder from the  MGRS and the 2dFGRS, respectively. 
  {\it  Upper right  panel:} Number  of  groups found  as function  of
  number of group members.  Solid histogram indicates results obtained
  from the 2dFGRS group catalogue, which is significantly inconsistent
  with  those  obtained  from  our  fiducial  MGRS.   To  remedy  this
  discrepancy,  either  clusters  have   to  have  a  relatively  high
  mass-to-light  ratio  of  $\sim  900  h  \MLsun$  (dashed  line)  or
  $\sigma_8  \simeq   0.7$  (dot-dashed  line)   {\it  Lower  panels:}
  Luminosity  of the  brightest galaxy  in each  group as  function of
  group mass.  Results  are shown for both the  2dFGRS (left) and MGRS
  (right)  group catalogues.   Errorbars  indicate 1-$\sigma$  scatter
  around the mean.  Solid  lines indicate two power-law relations, and
  are shown to facilitate a comparison.}
\label{fig:groups}
\end{figure}

Previous attempts to measure the kinematics of satellite galaxies have
mainly focussed  on isolated spiral  galaxies.  Using the  mock galaxy
redshift surveys described above we have investigated to what extent a
similar analysis  can be extended to  include a much  wider variety of
systems,  from isolated galaxies  to massive  groups and  clusters. In
particular, we used our MGRSs to optimize the host-satellite selection
criteria to  yield large  numbers of hosts  and satellites,  and small
fractions  of interlopers (satellites  not physically  associated with
the halo  of the host galaxy).   We found that  an iterative technique
with adaptive selection criteria  works best, allowing for an accurate
measurement of $\sigma_{\rm sat}(L_{\rm host})$; see [14] for details.

Applying our adaptive selection criteria  to the 2dFGRS yields a total
of  8132 host galaxies  and 12569  satellite galaxies.   The left-hand
panel  of Fig.~2 plots  the velocity  difference, $\Delta  V$, between
host and satellite galaxies as function of host luminosity. Notice the
increase of  the mean $\vert  \Delta V \vert$ with  increasing $L_{\rm
  host}$.   Fitting   the  distributions  $P(\Delta   V)$  of  various
luminosity bins with  a Gaussian plus constant (to  reproduce the true
satellites and interlopers, respectively), yields the relation between
$\sigma_{\rm sat}$  and $L_{\rm host}$  shown in the  right-hand panel
(solid dots  with errorbars). The gray area  indicates the expectation
values  obtained from  our CLF  where the  upper and  lower boundaries
outline the range of uncertainties due to the unknown second moment of
the  CLF (see  [14] for  details). Clearly,  the  satellite kinematics
obtained  from  the  2dFGRS  are  in excellent  agreement  with  these
predictions.   This provides  a  {\it dynamical}  confirmation of  the
average relation between mass and  light obtained from our purely {\it
  statistical} CLF formalism!

\section{Galaxy Groups}
\label{sec:groups}

We  recently  also  developed  a  halo-based  group  finder  that  can
successfully  assign galaxies  into groups  according to  their common
haloes [18].   The basic  idea behind our  group finder is  similar to
that  of the  host-satellite selection  criteria discussed  above.  In
short,  we start  with  an  assumed mass-to-light  ratio  to assign  a
tentative mass to each potential group.  This mass is used to estimate
the size and velocity dispersion of the underlying halo that hosts the
group,  which  in turn  is  used  to  determine group  membership  (in
redshift space).   This procedure is iterated  until group memberships
converge. Detailed tests with our MGRS show that this group finder (i)
is  $\gta 90$\%  complete in  terms of  group membership,  (ii) yields
interloper fractions  $\lta 20$\%,  and (iii) yields  group catalogues
that are  insensitive to the  initial assumption of  the mass-to-light
ratios.

Applying this group finder to the 2dFGRS yields a catalogue of $78708$
groups,  of which  7251  are  binaries, 2343  are  triplets, and  2502
contain four  members or  more. Group masses,  $M$, are  determined by
computing  the mean separation  between all  groups brighter  than the
group under  consideration and matching this with  the mean separation
between  dark  matter haloes  more  massive  than  $M$ (see  [20]  for
details).

The upper left-hand panel of  Fig.~3 plots the average number of group
members found  as function of  group mass.  The dashed  line indicates
the {\it true}  $\langle N \rangle(M)$ computed directly  from the CLF
used  to construct  the MGRS.   The  solid line  shows the  occupation
numbers  obtained from  the groups  selected from  the MGRS  using our
halo-based  group  finder.   Note  the excellent  agreement  with  the
`input' values,  indicating that  our group finder  is very  reliable. 
The  dotted line  indicates the  group occupation  statistics obtained
from our 2dFGRS  group catalogue.  Clearly, the CLF  predicts too many
galaxies per  group of a  given mass.  This  is also evident  from the
upper  right-hand panel,  where the  number  of groups  is plotted  as
function  of the  number of  group members.   The solid  histogram and
solid  line  indicate  the  results  for  the  2dFGRS  and  our  MGRS,
respectively.  Clearly, the MGRS, which  is based on our CLF, predicts
too many high-multiplicity groups.  To  remedy this, we need to either
set the average mass-to-light ratio on the scales of clusters to $\sim
900 h \MLsun$ (compared to $500  h \MLsun$ for our fiducial model), or
$\sigma_8$  has to be  reduced to  $\sim 0.7$.   This is  in excellent
agreement  with some  of our  previous results  based on  the pairwise
peculiar  velocity dispersions  [19] and  the abundances  of satellite
galaxies [15].

The lower panels of Fig.~3 plot the relation between the luminosity of
the brightest  (central) galaxy  in each group,  $L_c$, and  the group
(halo) mass,  $M$.  Results  are shown both  for groups in  the 2dFGRS
(left panel)  and for those in  our fiducial MGRS  (right panel).  The
mean $L_c$-$M$  relation is remarkably  similar for both  samples, and
well  described by  a broken  power-law with  $L_c\propto  M^{2/3}$ at
$M\la  10^{13}  h^{-1}M_{\odot}$  and  $L_c\propto M^{1/4}$  at  $M\ga
10^{13} h^{-1}M_{\odot}$.   At the low-mass end, this  is in excellent
agreement  with results  based  on galaxy-galaxy  weak lensing  (e.g.,
[17]).  At the massive end, $L_c$ only increases very slowly with halo
mass, indicating that  there must be a physical  process that prevents
the central galaxies in massive haloes from growing.

Finally, we want to emphasize that the groups can also be used to {\it
  directly}  measure the  halo ($=$  group) occupation  statistics. In
[22] we present some of these results, and show that the CLF of 2dFGRS
galaxies is perfectly consistent with  a Schechter form, contrary to a
recent claim by [23].

\section{Conclusions}

We  have  introduced  a  new  statistic,  the  conditional  luminosity
function $\Phi(L  \vert M)$, which  is ideally suited to  describe the
galaxy dark matter connection. The  CLF is well constrained using data
on the  clustering properties of  galaxies as function  of luminosity,
and yields a direct link between light and mass, which is in excellent
agreement with the kinematics of satellite galaxies.

Using the CLF, we have  identified two characteristic scales in galaxy
formation  (cf. [4]).   One at  $M \sim  10^{11} h^{-1}  \Msun$, where
galaxy formation is most efficient,  and one at $M \sim 10^{13} h^{-1}
\Msun$,  where  a transition  occurs  from  systems  dominated by  one
brightest, central galaxy to systems with several dominant galaxies of
comparable luminosity.

Using the CLF  we have also constructed detailed  mock galaxy redshift
surveys which  can be compared  on a one-to-one  basis to the  2dFGRS. 
The main  disagreements between  the MGRS and  the 2dFGRS  concern the
pairwise preculiar velocity dispersions  (see [19]), the abundances of
satellite  galaxies (see  [15]), and  the group  multiplicity function
(see Fig.~3).  These are all  reflections of one and the same problem:
the $\Lambda$CDM concordance cosmology predicts too many, large groups
and clusters,  unless the average  mass-to-light ratio of  clusters in
the photometric $b_J$-band is $\langle M/L \rangle_{\rm cl} \sim 900 h
\MLsun$ (with $\sigma_8=0.9$) or $\sigma_8 \simeq 0.75$ (with $\langle
M/L  \rangle_{\rm cl} \sim  500 h  \MLsun$). Recently,  these findings
have been confirmed  by a similar but independent  study of the galaxy
clustering properties in the SDSS [10].

\end{document}